# Modal Object Diagrams

Shahar Maoz*, Jan Oliver Ringert**, and Bernhard Rumpe

Software Engineering
RWTH Aachen University, Germany
http://www.se-rwth.de/

**Abstract.** While object diagrams (ODs) are widely used as a means to document object-oriented systems, they are expressively weak, as they are limited to describe specific possible snapshots of the system at hand. In this paper we introduce *modal object diagrams* (MODs), which extend the classical OD language with positive/negative and example/invariant modalities. The extended language allows the designer to specify not only positive example models but also negative examples, ones that the system should not allow, positive invariants, ones that all system's snapshots should include, and negative invariants, ones that no system snapshot is allowed to include. Moreover, as a primary application of the extended language we provide a formal verification technique that decides whether a given class diagram satisfies (i.e., models) a multi-modal object diagrams specification. In case of a negative answer, the technique outputs relevant counterexample object models, as applicable. The verification is based on a reduction to Alloy. The ideas are implemented in a prototype Eclipse plug-in. Examples show the usefulness of the extended language in specifying structural requirements of object-oriented systems in an intuitive yet expressive way.

> "...in the real world there are only objects.
> Classes exist only in our minds.", Nierstrasz [24]

## 1 Introduction

The language of object diagrams (ODs) is part of the UML standard and is supported by many academic and commercial software modeling tools (see, e.g., [7, 14, 23, 26, 32]). The semantics of an object diagram is simple: an object diagram describes a possible instantiation of the system under development – a single object model – a snapshot of the system's structure made of concrete object instances and the relations between them. However, while ODs are useful and intuitive means to present example instances of object-oriented systems in formal and semi-formal contexts, their expressive power is rather weak, and they have no additional usages beyond these simple presentations.

* S. Maoz acknowledges support from a postdoctoral Minerva Fellowship, funded by the German Federal Ministry for Education and Research.
** J.O. Ringert is supported by the DFG GK/1298 AlgoSyn.

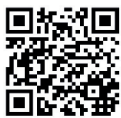

In this paper we introduce *modal object diagrams* (MODs), which extend the classical object diagram language with positive/negative and example/invariant modalities. The extended language allows to mark an object model not only as a *positive example*, describing a snapshot that the system should allow, but also as a *negative example*, which the system should not allow, as a *positive invariant*, which must be part of every snapshot of the system, or as a *negative invariant*, which must not be part of any snapshot of the system. Thus, the language supports the definition of very simple and intuitive yet expressively powerful specifications for the structure of the system to be; multi-modal specifications made of positive and negative examples and invariants. The syntax and semantics of modal object diagrams is formally defined in Sect. 3.

Moreover, as a primary application of the extended language, we consider a setup where a set of modal object diagrams is used as a specification that the system's class diagram (CD) should meet. The syntax of CDs is made of classes and the relationships between them, including inheritance and various associations. The semantics of CDs is given in terms of sets of objects and the relations between them. Thus, to support this application, we provide a formal verification technique that checks whether a CD – as a model of a system's structure defined by engineers – indeed satisfies (i.e., models) a multi-modal object diagrams specification. Given a CD and a multi-modal object diagram specification, the technique checks whether all positive examples in the specification are included in the CD's semantics, whether all negative examples are not included in the CD's semantics, whether all positive invariants are part of every object model in the CD's semantics, and whether all negative invariants are not part of any object model in the CD's semantics. In case of a negative answer, the technique outputs relevant counterexample object models, as applicable.

The verification technique is based on a transformation to Alloy [15]. Unlike previous works that consider the use of Alloy for the analysis of CDs (e.g., [1, 30]), the input for our transformation consists not only of a CD but also of an OD (or a set of ODs). Moreover, the transformation itself is different, as it follows a pragmatic approach: we are not suggesting a meta-model level framework for general transformations but instead focus on solving the concrete engineering problem we have at hand. The verification technique is described in Sect. 4.

In order to test and evaluate our work we have implemented and integrated it into a prototype Eclipse plug-in. The plug-in allows the engineer to edit MODs and CDs, and to verify a selected multi-modal MOD specification against a selected CD. Indeed, all examples shown in this paper have been verified by our plug-in. We describe the plug-in and the results of our experiments in Sect. 5.

The introduction of MODs and the associated verification technique suggest a stepwise design methodology. In early stages in the design process, MODs will most often be used by domain experts or system analysts, to describe possible snapshots of a system; in doing so, designers stipulate that the system should at least be able to exhibit the positive examples shown in the MODs. As the process matures, knowledge will become available about structures that should not be possible, so the initial set of positive example MODs could be refined with negative examples. Finally, in later stages, the analysts will be confident enough to define positive and negative invariant MODs. The verification technique we provide would aid the engineers in checking that their design indeed meets the concrete requirements defined by the MODs. We discuss this design process further in Sect. 7.4.

Object-oriented design constraints, similar to MOD specifications, can also be defined using the Object Constraint Language (OCL) [25]. Thus, one may consider using OCL instead of MOD or defining some combination of the two. We discuss the relation between OCL and MOD in Sect. 7.3.

The paper is organized as follows. The next section presents motivating examples for the use of multi-modal object diagram specifications. Sect. 3 formally defines the MOD language. Sect. 4 describes the verification problem and the technique to solve it. Sect. 5 presents our prototype implementation. Two extensions to the MOD language are discussed in Sect. 6. A discussion of limitations and advantages, a comparison with OCL, the use of MOD in the design process, and future work directions are presented in Sect. 7. Sect. 8 discusses related work and Sect. 9 concludes.

## 2 Examples

We start off with simple examples for multi-modal object diagram specifications, as they may be used during the design phase of a system. The examples are described semi-formally. The required formal definitions are given in Sect. 3.

### 2.1 Example I

Fig. 1 shows a specification $MS_1$ made of five MODs, prepared by a business analyst for a transportation services company. The specification includes three positive examples and two negative examples. *mod1.1* describes a car with a driver. *mod1.2* describes a car with two drivers. *mod1.3* describes a bus with a driver who has a manager. *mod1.1*, *mod1.2*, and *mod1.3* are all positive examples. *mod1.4* describes a negative example: a driver and a bus, not connected. Finally, *mod1.5* describes another negative example: a lone driver.

Given the specification $MS_1$, which was provided by the business analyst, the system's engineers have designed the class diagram $cd_1$ shown in Fig. 2. Note that the engineers have suggested to generalize `Car` and `Bus` using an abstract super class `Vehicle`. As this example is small, it is easy to see that $cd_1 \models MS_1$. The engineers have used our plug-in to verify this.

### 2.2 Example II

Following further investigation of the company's structure, the business analyst prepared a revised MOD specification $MS_2$, as shown in Fig. 3. The revised specification is made of four new MODs: a positive invariant, two negative invariants, and a positive example.

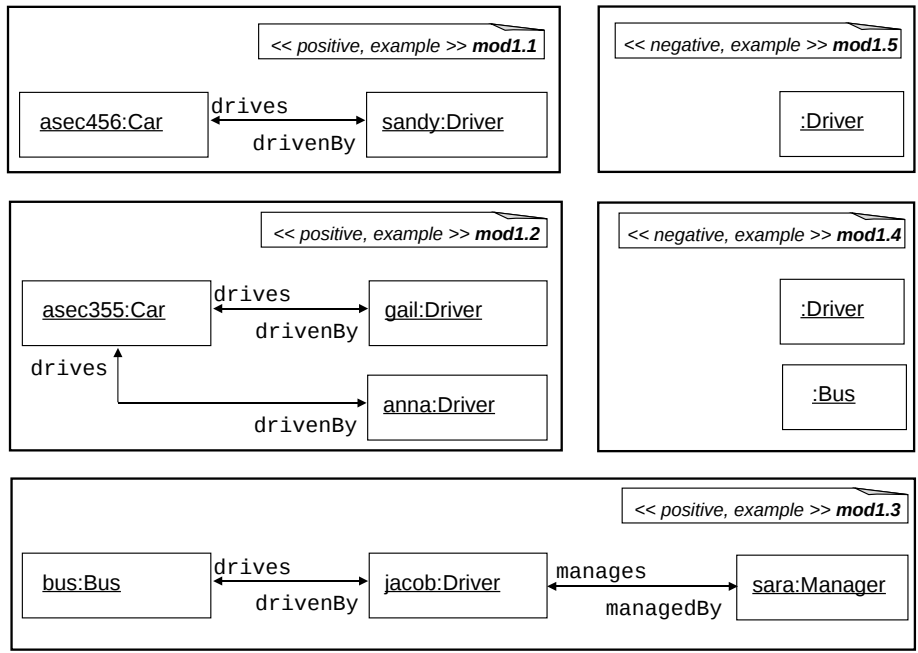

**Fig. 1.** The multi-modal MOD specification $MS_1$.

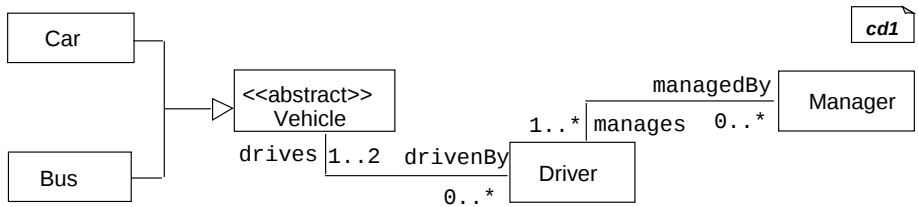

**Fig. 2.** $cd_1$, a class diagram prepared for the specification $MS_1$.

$mod2.1$ describes a positive invariant: every object model of the system must include at least one driver. $mod2.2$ describes a negative invariant: no object model of the system should include two managers. $mod2.3$ describes another negative invariant: a driver driving a bus, a car, and a sports car. Finally, $mod2.4$ describes a positive example: a manager managing an employee and a driver.

Given the $MS_2$ specification, the system's engineers have designed the CD $cd_2$ shown in Fig. 4. In the new CD, the engineers added a class Employee, and defined Driver and Manager to be its sub classes. They also set the class Manager to be a singleton, to support the negative invariant defined in $mod2.2$.

Unfortunately though, using our plug-in the engineers have found that $cd_2 \not\models MS_2$. First, $cd_2$ requires that every driver will drive at least one vehicle, but the driver in the positive example $mod2.4$ does not drive a vehicle. Second, $cd_2$ allows an object model consisting of a single manager that manages an employee that is not a driver. This contradicts the positive invariant *mod2.1* (the related counterexample model was suggested by our plug-in). In addition, $cd_2 \not\models MS_1$. For example, *mod1.1* shows a driver without a manager as a positive example, but according to $cd_2$ all object models should have exactly one manager (the same holds for *mod1.2*). Hence, the engineers should fix their CD or consult with the analyst on whether indeed a manager is required or not in every positive object model, and whether drivers should indeed drive at least one vehicle.

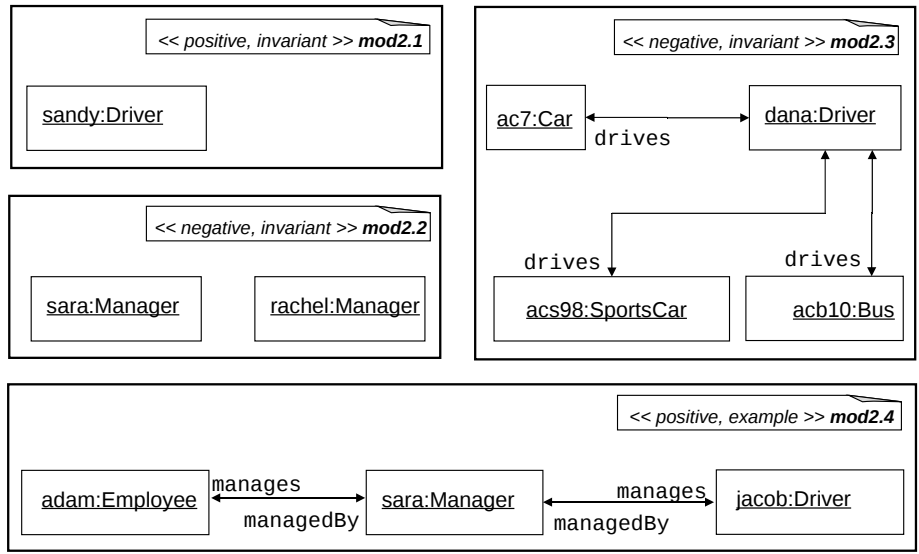

**Fig. 3.** The multi-modal MOD specification $MS_2$.

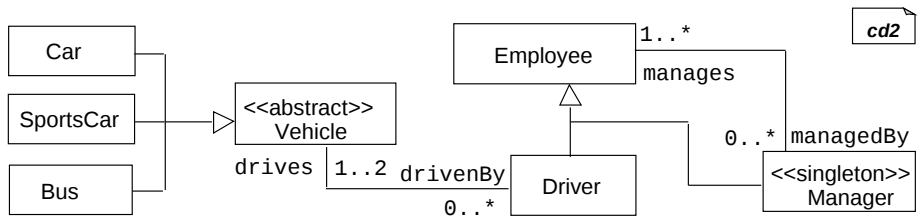

**Fig. 4.** $cd_2$, a class diagram prepared for the specification $MS_2$. Note, however, that $cd_2 \not\models MS_2$.

As the analysis progresses, the business analyst continues to learn about the system at hand and to provide the engineers with additional MODs. The engineers continue to design CDs that should meet the requirements set by the analyst and use our plug-in to check those CDs against the MODs that the analyst provides.

The next section presents the required formal definitions for modal object diagrams. We return to the above examples in the latter parts of the paper.

## 3 Modal Object Diagrams

We give an overview of the CD and OD languages used in our work, and continue with formal definitions of MODs and the relation $cd \models MS$ between a class diagram $cd$ and a multi-modal object diagram specification $MS$.

### 3.1 Class diagrams and object diagrams

As a concrete CD language we use the class diagrams of UML/P [27], a conceptually refined and simplified variant of UML designed for low-level design and implementation. Our semantics of CDs is based on [3,4,8] and is given in terms of sets of objects and relationships between these objects. More formally, the semantics is defined using three parts: a precise definition of the syntactic domain, i.e., the syntax of the modeling language CD and its context conditions (we use MontiCore [18,23] for this); a semantic domain - for us, a subset of the System Model (see [3,4]) OM, consisting of all finite object models; and a mapping $sem : CD \to \mathcal{P}(OM)$, which relates each syntactically well-formed CD to a set of constructs in the semantic domain OM. For a thorough and formal account of the semantics see [4].

For example, the semantics of $cd_1$ shown in Fig. 2 includes all object models where all drivers drive one or two vehicles, all vehicles are driven by zero or more drivers, there are no vehicles that are not cars or buses but there may be cars and buses, every driver has zero or more managers, and every manager manages at least one driver. Note that the empty object model, which is an object model with no objects at all, is in the semantics of $cd_1$. In addition, note that the semantics of $cd_1$ consists of an infinite number of object models.

As another example, the semantics of $cd_2$ shown in Fig. 4 includes all object models where all drivers drive one or two vehicles, all vehicles are driven by zero or more drivers, there are no vehicles that are not cars, buses, or sports cars, but there may be cars and buses and sports cars, every employee has zero or more managers, every driver is an employee, every manager is an employee, every manager manages at least one employee, and there is exactly one manager. The empty object model is not in the semantics of $cd_2$ because every object model in the semantics of $cd_2$ must include exactly one manager. Also, as in the semantics of $cd_1$, the semantics of $cd_2$ consists of an infinite number of object models.

Note that we use a *complete* interpretation for CDs (see [27] ch. 3.4), roughly meaning that 'whatever is not in the CD, should indeed not be present in the object model'. In particular, we assume that the list of attributes of each class is complete, e.g., a `driver` object with an `address` and a `salary` is not considered as part of the semantics of a `Driver` class with an `address` only. Also, the list of classes in the CD is considered complete, in the sense that its object models cannot include objects of classes not explicitly mentioned in the CD.

Also note that object names may be used in the OD for convenience, but they have no semantic meaning, i.e., the object name is not interpreted as an attribute `name`. Thus, e.g., an object `sara:Driver` has the same semantics as an object `dan:Driver` or an unnamed object `:Driver`.

The CD language constructs we support include generalization (inheritance), interface implementation, abstract and singleton classes, class attributes, uni- and bi-directional associations with multiplicities, enumerations, aggregation, and composition.

A note about notation: object diagrams refer to concrete syntactical expressions and object models refer to elements in the semantic domain. Still, the mapping from the abstract syntax to the semantic domain is in this case one-to-one, so we use OD and OM interchangeably. For example in the definition below we write $\forall pe \in MS.PE : pe \in sem(cd)$, although $MS.PE$ is a set of (modal) object diagrams while $sem(cd)$ is a set of object models. A more strict (yet inconvenient) notation should use $sem(pe) \in sem(cd)$.

### 3.2 Defining modal object diagrams

We are now ready to present modal object diagrams, multi-modal object diagram specifications, and their relation to class diagrams.

**Definition 1 (modal object diagram (MOD)).** *A modal object diagram is a structure $mod = \langle od, p, q \rangle$ where $od \in OD$ is an object diagram, $p \in \{positive, negative\}$, and $q \in \{example, invariant\}$.*

Syntactically, we use stereotypes to denote the positive/negative and example/invariant modalities. Alternative syntactic means may be suggested, e.g., the use of dashed-line boxes in example MODs vs. solid-line boxes in invariant MODs.

**Definition 2 (multi-modal object diagram specification).** *A multi-modal object diagram specification is a set of MODs. Given a specification MS, the set of positive example MODs in MS is denoted MS.PE, the set of negative example MODs in MS is denoted MS.NE, the set of positive invariant MODs in MS is denoted MS.PI, and the set of negative invariant MODs in MS is denoted MS.NI. Any of these sets may be empty.*

We define the satisfaction relation between a CD and a multi-modal object diagram specification. Below we use $om_1 \subseteq om_2$ to note that all objects and links appearing in $om_1$ appear also in $om_2$.

**Definition 3 ($cd \models MS$).** *A class diagram cd satisfies a multi-modal object diagram specification MS, denoted $cd \models MS$, iff*

1. $\forall pe \in MS.PE : pe \in sem(cd)$;
2. $\forall ne \in MS.NE : ne \notin sem(cd)$;
3. $\forall pi \in MS.PI, \forall om \in sem(cd) : pi \subseteq om$;

4. $\forall ni \in MS.NI, \forall om \in sem(cd) : ni \nsubseteq om$.

Note that the definition above uses a *complete*, rather than a *partial*, interpretation of positive example MODs. That is, it considers each OD to describe a complete object model rather than a partial one. We discuss an alternative partial semantics variant in Sect. 6.1.

Finally, since our verification technique, as described in the next section, is based on a transformation to Alloy [15], we need a bounded variant of the satisfaction relation. Below we use $|om|$ to note the maximal number of objects per class in $om$ (objects of subclasses are counted also as objects of their super classes). Note that a bound needs to be applied only to invariants, because for example MODs, the MOD itself determines the size of the problem.

**Definition 4 ($cd \models_k MS$).** *A class diagram $cd$ satisfies a multi-modal object diagram specification MS modulo a bound $k > 0$, denoted $cd \models_k MS$, iff*

1. $\forall pe \in MS.PE : pe \in sem(cd)$;
2. $\forall ne \in MS.NE : ne \notin sem(cd)$;
3. $\forall pi \in MS.PI, \forall om \in sem(cd) \text{ s.t. } |om| \leq k : pi \subseteq om$;
4. $\forall ni \in MS.NI, \forall om \in sem(cd) \text{ s.t. } |om| \leq k : ni \nsubseteq om$.

## 4 Verifying a CD Against an MOD Specification

### 4.1 Problem definition

The verification problem definition is as follows: given a multi-modal object diagram specification $MS$, a class diagram $cd$, and a bound $k$, check whether $cd \models_k MS$. Moreover, in case of a negative answer, provide relevant counterexample object models, as applicable for the negative and positive invariants at hand: for each unsatisfied $pi \in MS.PI$, provide $om \in sem(cd)$ s.t. $pi \nsubseteq om$; for each unsatisfied $ni \in MS.NI$, provide $om \in sem(cd)$ s.t. $ni \subseteq om$.

Our solution is based on a transformation to Alloy [15].

### 4.2 A brief overview of Alloy

Alloy is a textual modeling language based on relational first-order logic. An Alloy module consists of a number of signature declarations, fields, facts and predicates. The basic entities in Alloy are atoms. Each signature denotes a set of atoms. Each field belongs to a signature and represents a relation between two or more signatures. Such relations are interpreted as sets of tuples of atoms. Facts are statements that define constraints on the elements of the module. Predicates are parametrized constraints, which can be included in other predicates or facts.

Alloy Analyzer is a fully automated constraint solver for Alloy modules. The analysis is achieved by an automated translation of the module into a Boolean expression, which is analyzed by SAT solvers embedded within the Analyzer. The analysis is based on an exhaustive search for instances of the module. The search space is bounded by a user-specified scope, a positive integer that limits the number of atoms for each signature in an instance of the system that the solver analyzes.

The Analyzer checks for the validity of user-specified assertions. If an instance that violates the assertion is found within the scope, the assertion is not valid. If no instance is found, the assertion might be invalid in a larger scope. Used in the opposite way, one can look for instances of user-specified predicates. If the predicate is satisfiable within the given scope, the Analyzer finds an instance that proves it. If not, the predicate may be satisfiable in a larger scope. Sect. 7 discusses the advantages and limitations of using Alloy for our problem.

### 4.3 Solution by transformation to Alloy

The transformation consists of three parts: handling the CD, handling each of the MODs, and generating of Alloy run commands. The complete transformation details are given in [19]. Here we give an overview of the transformation, using generated Alloy code taken from some of the examples shown earlier in Sect. 2.

**Handling the CD** The input CD is transformed into a set of Alloy signatures, functions, and facts. Each class is transformed into an Alloy signature of a corresponding name, with fields defined according to the associations given in the CD. Alloy functions are defined in order to specify sets of objects of specific concrete classes, taking into account the information about inheritance hierarchy defined in the CD. Finally, Alloy facts are defined to express the types and multiplicities involved in the associations defined in the CD.

For example, the generated Alloy signatures, functions, and facts for the class diagram $cd_2$ of Fig. 4, are shown in Listings 1.1, 1.2, and 1.3.

Listings 1.1 shows the Alloy signatures for all the classes defined in the CD, with fields defined according to their associations. For example, see the `managedBy` field defined in line 2 for the signature `Employee`. The keyword `extends` is used to model class inheritance (see, e.g., lines 13-15, where the three sub classes of `Vehicle` are defined). Alloy's `one` keyword is used to model the singleton requirement specified by the `singleton` stereotype in the CD (see line 7). Finally, Alloy's `abstract` keyword is used to model the abstract requirement specified by the `abstract` keyword in the CD (see line 10).

Listings 1.2 shows the Alloy functions that specify sets of objects of concrete classes. Each function specifies a single set consisting of all objects of its corresponding class only, that is, without objects of its sub classes. For example, in line 1, the function `EmployeeOnly` is defined as the set consisting of all employees that are not drivers or managers.

Listings 1.3 shows the types and multiplicities of the associations defined in $cd_2$. The `VehicleIsAbstract` fact specifies that the set `VehicleOnly` includes no elements. The symmetry of the bi-directional association between `Driver` and `Vehicle` is specified by requiring that the restriction of `Driver` to the field `drives` (as a relation) is the inverse of the restriction of `Vehicle` to the field

```
1 sig Employee  {
2   managedBy: set Manager
3 }
4 sig Driver extends Employee {
5   drives: set Vehicle
6 }
7 one sig Manager extends Employee {
8   manages: set Employee
9 }
10 abstract sig Vehicle  {
11   drivenBy: set Driver
12 }
13 sig Car extends Vehicle { }
14 sig Bus extends Vehicle { }
15 sig SportsCar extends Vehicle { }
```

**Listing 1.1.** Generated Alloy signatures for $cd_2$.

```
1 fun EmployeeOnly: set univ {Employee-(Driver + Manager)}
2 fun DriverOnly: set univ {Driver}
3 fun ManagerOnly: set univ {Manager}
4 fun VehicleOnly: set univ {Vehicle-(Car+Bus+SportsCar)}
5 fun CarOnly: set univ {Car}
6 fun BusOnly: set univ {Bus}
7 fun SportsCarOnly: set univ {SportsCar}
```

**Listing 1.2.** Generated Alloy functions for $cd_2$.

drivenBy. Note that by definition this applies to all the sub classes of Vehicle too. The multiplicity constraints of the association between Driver and Vehicle are specified by limiting the size of the relevant sets referenced by the corresponding signatures' fields. The last two facts specify the symmetry and the multiplicities constraints for the association between Employee and Manager in a similar way.

**Handling each of the MODs** Each input MOD is transformed into a predicate. The predicate consists of a conjunction of a number of parts. First, all objects in the diagram are listed (anonymous objects are given a random unique name). Second, the concrete types and number of occurrences is specified, making sure that super classes are not handled as sub classes. Finally, the links between the objects are specified using field assignments.

Listing 1.4 shows the generated predicate for the positive example MOD $mod2.4$ of Fig. 3. Lines 3-4 list the names of the objects and their types by declaring corresponding variables. Lines 6-7 specify the concrete classes these objects belong to. Line 8 defines the number of objects of each class (e.g., if there were two or more managers, as in $mod2.2$, this would include the statement #{sara, rachel} == 2, to make sure that each variable references a distinct

```
1  fact VehicleIsAbstract { # VehicleOnly == 0 }
2  fact Asso_Driver_drives_drivenBy_Vehicle_symmetry {
3    Driver <: drives = ~((Vehicle <: drivenBy))
4  }
5  fact Asso_Driver_drivenBy_drives_Vehicle_Mult {
6    all var: Driver | # var.drives >= 1 && # var.drives <= 2
7    all var: Vehicle | # var.drivenBy >= 0
8  }
9  fact Asso_Employee_managedBy_manages_Manager_symmetry {
10   Employee <: managedBy = ~((Manager <: manages))
11 }
12 fact Asso_Employee_manages_managedBy_Manager_Mult {
13   all var: Employee | # var.managedBy >= 0
14   all var: Manager | # var.manages >= 1
15 }
```

**Listing 1.3.** Generated Alloy facts for $cd_2$.

```
1  pred checkFull {
2    // all objects in our OD
3    some adam: Employee | some jacob: Driver |
4      some sara: Manager |
5    // make sure a superclass is not handled as a subclass
6    adam in EmployeeOnly and jacob in DriverOnly
7      and sara in ManagerOnly
8    and # {adam} == 1 and # {jacob} == 1 and # {sara} == 1
9    // define universe
10   and univ = {adam + jacob + sara + Int}
11
12   // links between them
13   and sara.manages = {jacob + adam}
14   and jacob.managedBy = {sara}
15   and adam.managedBy = {sara}
16   and jacob.drives = none
17 }
```

**Listing 1.4.** Generated predicate for the positive example $mod2.4$. If this was an invariant (positive or negative), the predicate would have been named `checkPart` and the conjunct defining the universe (line 10) would have been omitted.

object). Line 10 specifies that the universe of objects for Alloy is exactly the set of objects listed in the MOD (this statement is omitted for invariant MODs, see below). Finally, lines 13-16 specify the concrete associations between the objects.

The difference between example MODs and invariant MODs (whether positive or negative) is manifested in the transformation as follows. For example MODs, the predicate includes an additional conjunct, which defines the universe for Alloy as exactly the set of objects listed in the diagram. For invariant MODs,

```
1 run checkFull for 6 but exactly 1 Driver,
2   exactly 3 Employee, exactly 1 Manager
```

**Listing 1.5.** Generated Alloy run command for the positive example *mod2.4*.

in contrast, this conjunct is not added, as the universe for Alloy is allowed to include more objects.

As described above, Listing 1.4 shows the generated predicate for the positive example MOD *mod2.4* of Fig. 3. If the MOD was not an example but an invariant MOD (positive or negative), the generated predicate would have been named `checkPart` instead of `checkFull` and would not have included the conjunct defining the universe. Lines 13-16 would also have changed, to reflect that additional links may exist.

**Generating run commands for Alloy** Finally, we generate a set of run commands for Alloy, each corresponding to one of the MODs in the specification.

For positive or negative example MODs, the run command simply runs the corresponding generated predicate with exact per-class scopes taken from the MOD itself (the computation of the per-class scope takes super classes into consideration too). As an example, the run command for the positive example MOD *mod*2.4 is given in Listing 1.5. The same run command is used to check for a negative example; the only difference between negative and positive examples is in the interpretation of the result that is provided by Alloy.

Checking for a positive invariant is done by asserting the generated `checkPart` predicate. That is, the generated command checks an Alloy assertion named `checkInvariant`, which includes the `checkPart` predicate. The scope for this check cannot be taken from the input MOD and is defined by the user. As an example, the check command for a positive invariant MOD, with a user-defined scope of 6, is given in Listing 1.6. Checking for a negative invariant is done using the `checkPart` predicate, with a user-defined scope.

## 5 Implementation and Evaluation

In order to test and evaluate the MOD language and the verification technique we have implemented a prototype Eclipse plug-in that allows the engineer to edit MODs and CDs and to verify a selected CD against an MOD specification.

```
1 assert checkInvariant {
2   checkPart
3 }
4
5 check checkInvariant for 6
```

**Listing 1.6.** Generated Alloy assert statement and check command for a positive invariant MOD, with a user defined scope of 6.

The prototype implementation, examples, and related materials are available from [22].

The plug-in includes a textual editor for MODs and CDs. The editor was generated using MontiCore [23] grammars (including a parser, syntax highlighting, outline view etc.), and the addition of the modalities to the ODs is done using stereotypes. When the user selects a number of MODs and a CD, she can execute the verification. The transformation to Alloy is implemented using templates written in FreeMarker [9] and the execution of the generated module run commands is done using Alloy's APIs (the analysis is fully automated so the engineer does not need to see the generated Alloy code). Several parameters may be selected, e.g., the SAT solver and the scope that Alloy will use in the analysis.

The results of the verification process are shown in a hierarchical table (an Eclipse view), displaying which MODs are modeled by the CD and which ones are not. When the engineer clicks the name of an invariant MOD that is not modeled by the selected CD, if any, the plug-in displays a relevant generated counterexample OD in the main editor pane of the IDE.

### 5.1 Example results

We have used the plug-in to examine several MOD specifications and related class diagrams, including the examples shown in Sect. 2. Fig. 5 shows a screen capture of the eclipse IDE, displaying several CD and MOD files on the explorer view on the left (files with extensions `.cd` and `.od`), the summary results of a verification process at the bottom of the screen, and a generated counterexample OD in the main view (an outline of the OD is shown on the right).

Specifically, the figure shows the results of verifying $cd_2$ against the four MODs of the multi-modal specification $MS_2$ (shown earlier in figures 4 and 3). As expected, the results table shows two failures and two passes. Indeed, $cd_2 \not\models \{mod2.1\}$ and $cd_2 \not\models \{mod2.4\}$. The counterexample shown in the main view relates to the positive invariant $mod2.1$: it shows an object diagram that consists of an employee, a manager, and two cars, where the manager manages herself and the other employee. This object model is in the semantics of $cd_2$ but it does not include a driver. Thus, it proves that the positive invariant $mod2.1$ is not satisfied by $cd_2$.

As the example shows, the counterexample found by Alloy is not necessarily the smallest possible counterexample (within the user-specified scope). This points to a limitation in our technique. In the future, it may be worthwhile to develop a technique that produces minimal counterexamples according to some minimality criteria (e.g., number of classes, total number of objects, etc.).

### 5.2 Performance results

Table 1 shows performance results from our experiments. Experiments were done using Alloy version 4.1.10 with SAT4J [28], on a regular laptop computer, Intel

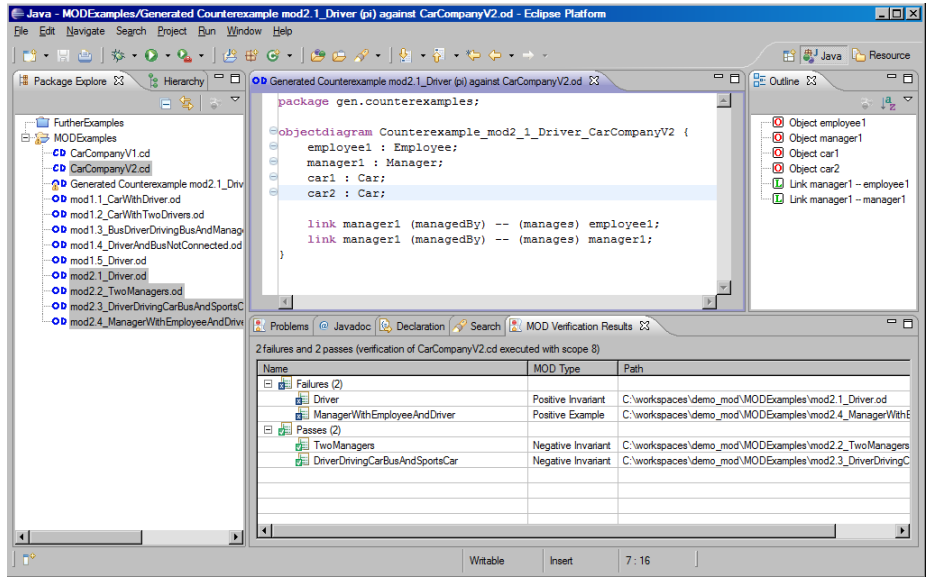

**Fig. 5.** A screen capture from Eclipse, displaying several CD and MOD files on the explorer view on the left (files with extensions .cd and .od), the summary results of a verification process at the bottom of the screen, and a generated counterexample OD in the main view (an outline of the OD is shown on the right). Specifically, the figure shows the results of verifying $cd_2$ against the four MODs of the multi-modal specification $MS_2$, with two failures and two passes. The counterexample relates to the verification of the positive invariant *mod2.1*.

Dual Core CPU, 2.8 GHz, with 4 GB RAM, running Windows Vista. The CDs and MODs are the ones presented in Sect. 2.

For each CD and MOD, the table shows whether the verification passed or failed (i.e., whether the CD satisfies the MOD or not), some details on the SAT formula that Alloy generated (number of variables etc.), and the total time it took to run the verification (constructing the formula + solving it), in milliseconds. The column titled `Scope` reports on the scope used in the verification: as explained in Sect. 4.3, for invariants MODs we use a user-defined scope; for example MODs, specific per-class scopes are taken from the MOD itself.

Interestingly, for some of the MODs, the generated Alloy formula was empty (had zero variables). That is, for these MODs, Alloy was able to determine the result without using the SAT solver. This happens when the generated Alloy module is very simple, e.g., when the `checkPart` predicate includes an immediate violation (contradiction) of one of the facts.

The verification of example MODs is, in general, simpler and faster, as it takes per-class scopes from the MOD at hand and its solution space is relatively small. The verification of invariant MODs, in contrast, is more complicated, and its solution space and performance depends on the user defined scope.

| CD | MOD | Modalities | Scope | Result | Vars/primary vars/clauses | Time (ms) |
|---|---|---|---|---|---|---|
| $cd_1$ | $mod1.1$ | PE | by OD | pass | 141 / 19 / 209 | 20 + 7 |
| $cd_1$ | $mod1.2$ | PE | by OD | pass | 371 / 34 / 581 | 18 + 8 |
| $cd_1$ | $mod1.3$ | PE | by OD | pass | 37 / 7 / 57 | 5 + 5 |
| $cd_1$ | $mod1.4$ | NE | by OD | pass | 132 / 19 / 192 | 6 + 0 |
| $cd_1$ | $mod1.5$ | NE | by OD | pass | 276 / 36 / 449 | 6 + 1 |
| $cd_2$ | $mod2.1$ | PI | 8 | fail | 2107 / 166 / 4360 | 17 + 10 |
| $cd_2$ | $mod2.2$ | NI | 8 | pass | 0 / 0 / 0 | 7 + 0 |
| $cd_2$ | $mod2.3$ | NI | 8 | pass | 2794 / 197 / 5867 | 17 + 7 |
| $cd_2$ | $mod2.4$ | PE | by OD | fail | 263 / 36 / 466 | 4 + 0 |
| $cd_2$ | $mod1.1$ | PE | by OD | fail | 0 / 0 / 0 | 5 + 0 |
| $cd_2$ | $mod1.2$ | PE | by OD | fail | 0 / 0 / 0 | 3 + 0 |
| $cd_2$ | $mod1.3$ | PE | by OD | pass | 49 / 9 / 71 | 4 + 11 |
| $cd_2$ | $mod1.4$ | NE | by OD | pass | 0 / 0 / 0 | 2 + 0 |
| $cd_2$ | $mod1.5$ | NE | by OD | pass | 0 / 0 / 0 | 4 + 0 |

**Table 1.** Results from experimenting with the verification of example MODs and CDs.

The performance results in Table 1 show that for relatively small models, MOD/CD verification runs very fast. However, we do have other examples, not shown here, where MOD verification uses many more variables and clauses and takes much more time to compute. Given these results, in the future, we plan to develop heuristics to improve the scalability of invariant MOD verification, using, e.g., abstraction / refinement techniques, decomposition for early detection of independent sub models, etc. See the short discussions in sections 7.1 and 7.2.

## 6 Extensions

We present and discuss two extensions to the basic MOD language, inspired by [27]: partial vs. complete positive examples and parametrized ODs.

### 6.1 Partial vs. complete positive examples

We distinguish between partial and complete positive examples. Roughly, a partial positive example object diagram specifies an object model that should be extensible to a positive example object model.

Recall the examples discussed in Sect. 2. There, we saw that $cd_2 \not\models MS_1$ because the positive examples $mod1.1$ and $mod1.2$ did not include a manager. Using the distinction between partial and complete positive examples, the analyst can specify that these MODs are partial positive examples and not complete ones. Doing so will make $cd_2$ satisfy $MS_1$.

Syntactically, we specify that an OD is partial using a stereotype `partial`. The semantics for partial positive examples is formally defined as follows:

**Definition 5** ($cd \models ppe$). *Given a class diagram cd and a partial positive example object diagram $ppe = \langle od, partial, positive, example \rangle$, we say that cd satisfies ppe, denoted $cd \models ppe$, iff $\exists od' \in sem(cd)$ s.t. $od \subseteq od'$.*

Updating Definitions 2, 3, and 4 from Sect. 3 to include partial positive examples is straightforward. The combination of `partial` with modalities other than positive examples is considered syntactically incorrect and its semantics is undefined (since invariant MODs already have a partial interpretation).

Note that the verification technique we described in Sect. 4 already supports MOD specifications with partial positive examples. Specifically, recall the `checkPart` predicate, which we use in the computation of invariants within an assert statement. Running this predicate without an assert provides the required verification for partial positive examples.

### 6.2 Parametrized object diagrams

We extend the classical object diagram language with parameters, which may be used for attribute values or for object types. For example, instead of assigning a specific value to an attribute, the designer may assign it a parameter, and then define the set or range of values this parameter may take. Multiple parameters may be used in a single object diagram and the same parameter may appear more than once in a diagram. The semantics of parametric object diagrams is a natural extension of the classical semantics: it consists of the set of object models obtained by creating a set of non-parametric copies of the diagram, where in each copy different (combination of) values are assigned to the parameters.

For example, to specify that a driver's experience level can be either novice, regular, or expert, only a single (positive example) object diagram needs to be drawn, showing a driver whose `experience` attribute equals the OD parameter `level`, and $level \in \{novice, regular, expert\}$. Thus, the parametric extension allows the designer to create succinct object diagram specifications.

The combination of parametric object diagrams and modal object diagrams yields a powerful specification language. We give an example below.

Fig. 6 shows an MOD specification $MS_3$, made of three parametric MODs. *mod3.1* is a negative invariant: it specifies that a driver cannot have an age lower than 16 (more formally, that any object model of the system should not include a driver whose age is between 1 and 15). *mod3.2* is a positive example: it specifies that a driver can drive a car, a sports car, or a bus (more formally, that a driver driving a car, a driver driving a sports car, and a driver driving a bus, are all positive examples of object models of the system). Finally, *mod3.3* is a negative example: it specifies that a driver who has novice or regular level of experience, is not allowed to drive a sports car that has medium or high engine power.

The verification technique we presented can be extended to support the parametric extension. The corresponding CDs may need to be enhanced with OCL constraints, which may get rather complicated, so the analysts and engineers should agree on the level of detail they want the specification and CDs to use.

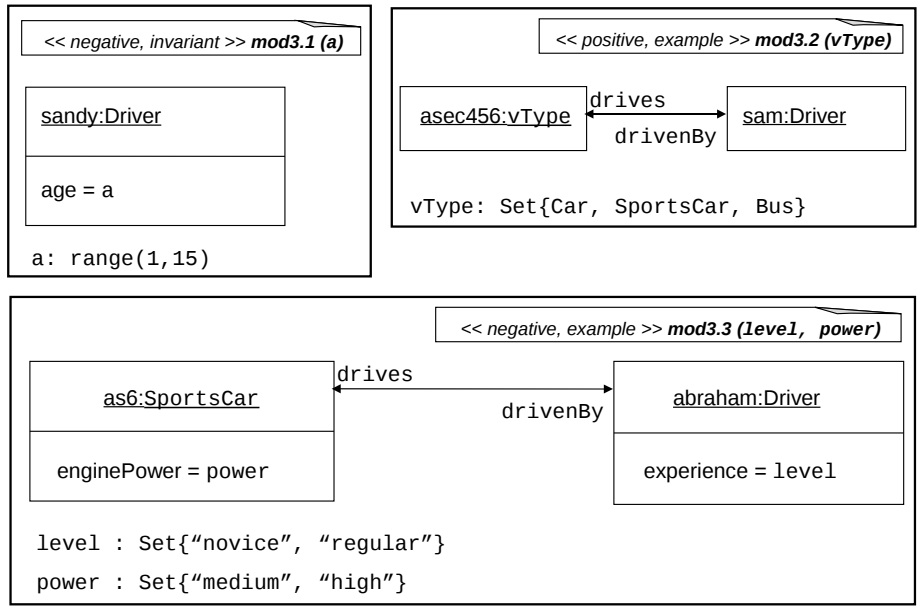

**Fig. 6.** The multi-modal parametrized MOD specification $MS_3$.

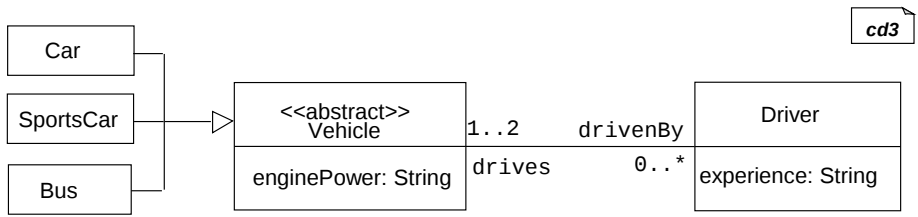

**Fig. 7.** A class diagram for the parametrized MOD specification $MS_3$.

## 7 Discussion and Future Work

We discuss advanced topics related to our work, its advantages and limitations, and related future work directions. These include a discussion of complexity and performance, the bounded scope limitation, the relationship between MOD and OCL, the use of MOD in the design process, and the problem of synthesizing a CD from an MOD specification.

### 7.1 Complexity and performance

The transformation of the CD and MODs to Alloy is linear in the size of the input diagrams. It requires only a constant number of iterations over the diagrams' syntax and the construction of constant number of linear size 'symbol tables'.

The computation by Alloy using a SAT solver, may be exponential in the size of the input diagrams. For example MODs, the solution space is relatively small and depends on the number of objects defined in the OD. In the case of invariant MODs, the solution space depends on the user-defined scope $k$. Although the two problems are rather different, we use a unified approach that solves both.

As discussed in Sect. 5, experience shows that our technique works very fast for relatively small models. Indeed, most works dealing with reasoning about CDs use rather small CDs in their experiments (see, e.g., [31, 33]). Still, as future work, to make MOD verification practical for real-world projects, it would be necessary to develop heuristics that may accelerate the performance of Alloy in verifying larger MODs, experiment with the different SAT solvers supported by Alloy, or define a direct translation into SAT (as was suggested in [31]).

### 7.2 The bounded scope limitation

The verification of positive and negative example object diagrams is sound and complete. The required scopes are taken from the example object diagrams themselves and so the answer is not only sound but complete.

The verification of positive and negative invariant object diagrams is however bounded by the user-defined scope. Specifically, it may be the case that for some $cd$ and $MS$, $cd \not\models MS$ but $cd \models_k MS$ for some given $k$. A simple concrete example is as follows. Consider a CD $cd_a$ consisting of two classes, $C_1$ and $C_2$, and an association of multiplicity of exactly 1 to 5 between $C_1$ and $C_2$. Assume a a negative invariant object diagram $ni$ consisting of two instances of $C_1$ and no instances of $C_2$. Clearly, $cd_a \not\models \{ni\}$ but $cd_a \models_k \{ni\}$ for any $k < 10$.

The bounded version $cd \models_k MS$ is thus indeed strictly weaker than the general version $cd \models MS$. Our use of Alloy is sound and complete for the bounded version but is neither sound nor complete for the general version.

To conclude, the use of Alloy, and consequently the encoding of our verification problem as an instance of SAT, carries the significant price of bounded analysis. Nevertheless, we adapt the small scope hypothesis of [15] to our domain, and suggest that in many cases, although the models involved may be large, counterexamples for their unsatisfaction could be rather small.

As future work, heuristics may be developed to make automatic or semi-automatic informed guesses about suitable scopes that could reduce given problems into equivalent smaller ones where lower scopes are 'good enough', or to identify cases where one could automatically prove that a higher scope will not change the analysis results.

### 7.3 MOD and OCL

The Object Constraint Language (OCL) [25] is a declarative language for describing rules that apply to Unified Modeling Language (UML) models (and, more generally, to any Meta-Object Facility (MOF) meta-model). OCL is based on first-order predicate logic. As MODs specify constraints on object-oriented models too, discussing the relationship between MOD and OCL is worthwhile.

```
package cd01
    context Manager
    inv inv01: not
        (Manager.allInstances()->
            exists( sara:Manager | Manager.allInstances()->
                exists( rachel:Manager |
                    -- the two objects are distinct
                    not(sara=rachel) and
                    -- objects are of the specified types
                    -- and not of any of their sub types
                    sara.oclIsTypeOf(Manager) and
                    rachel.oclIsTypeOf(Manager) and
                    -- sara and rachel do not manage anyone
                    sara.manages.asSet()->size()=0 and
                    rachel.manages.asSet()->size()=0 and
                    -- sara and rachel not managed by anyone
                    sara.managedBy.asSet()->size()=0 and
                    rachel.managedBy.asSet()->size()=0
                )
            )
        )
endpackage
```

**Fig. 8.** An OCL representation of MOD *mod2.2* from Fig. 3.

OCL is interpreted in the context of a UML diagram and is limited to specifying invariants, i.e., constraints that hold for all its instantiations. Thus, given that a CD context is provided, invariant MODs, positive and negative, can be specified using OCL. Moreover, negative example MODs can also be specified in OCL, by specifying a negative invariant that constrain also the set of all instances (the universe) to the set of existing instances listed in the MOD. Positive example MODs cannot be specified in OCL.

As a small example, recall *mod2.2*, which shows a negative invariant MOD. This MOD is semantically equivalent to the OCL code shown in Fig. 8 (to edit the OCL code that we show we used the Dresden OCL Eclipse plug-in [6]). If *mod2.2* was a negative example, the OCL code in Fig. 8 should have been extended, to specify, inside the outermost negative clause, that the total number of managers is two. Furthermore, every other class from the CD should be listed, to specify that the size of its instances set is zero. For comparison purposes, Fig. 9 shows the textual representation of MOD *mod2.2* in our object diagram language (the language grammar is defined in MontiCore [18, 23]).

These examples demonstrate that although it is impossible to specify positive example MODs in OCL, it is formally possible to specify invariant and negative example MODs using OCL. Yet, it is clearly very inconvenient, because manual writing of such OCL statements is obviously technically difficult and error prone. Thus, we chose to introduce MODs due to their readable and

```
package test.examples;

<<negative, invariant>> objectdiagram TwoManagers {
    sara : Manager;
    rachel : Manager;
}
```

**Fig. 9.** The textual representation of MOD *mod2.2* from Fig. 3, as used in our work.

succinct representation, which makes them usable and attractive, using either textual or visual concrete syntax, not only for software engineers but also for non-SE specialists such as business analysts or other domain experts. On top of classical ODs, MODs make the notion of modality explicit; they integrate the intuitive concrete representation of the OD language with a limited set of predefined natural modalities.

Finally, as OCL is much richer than MOD in the kinds of invariants it can specify in the context of a given diagram, it may be interesting to follow [27] and define a combination of OD (MOD) and OCL. We discuss this combination in the related work section.

### 7.4 Using MODs in the design process

Just like classical ODs, MODs are simple and intuitive to define, since the addition of modalities does not change the basic syntax and semantics of describing a single concrete instance. In particular, ODs are much simpler than CDs, as they do not show inheritance and interface implementation relations. Moreover, CDs are made of abstract entities – classes – which do not 'exist' in the 'real world'. As Oscar Nierstrasz puts it, "Classes exist only in our minds" [24]. ODs, in contrast, are made of concrete entities – objects – which indeed 'exist', both in the real world and in the systems we build, when they run.

The introduction of the MOD language suggests a stepwise design methodology. In early stages in the design process, MODs will most often be used by domain experts and analysts to describe possible snapshots of a system. In doing so, they would stipulate that the system should at least be able to exhibit the examples shown in the MODs. That is, only positive example MODs will be used in the early stages of the design process. As the process matures, knowledge will become available about structures that should not be possible, so the initial set of positive example MODs could be refined with negative examples. Finally, in later stages, analysts will be confident enough to define positive and negative invariant MODs.

The MOD language and this design process are inspired by an analogous design process for behavioral specifications. There, domain experts may provide positive example execution traces, which the system should allow, negative example traces, which the system should not allow, invariant traces, which all system executions should include, and negative invariant traces, which no ex-ecution that allows them can be extended to an accepted one. Then, software engineers are responsible for designing a state-machine that will satisfy these multi-modal trace requirements, and model-checking techniques can be used to verify them. Such concrete multi-modal traces can be specified, e.g., using live sequence charts (LSC) [5, 12] (see the related work section).

To conclude, we believe that the MOD language can be used not only by software engineers but also by domain experts and analysts, in particular during early requirements phases of object-oriented systems. Moreover, the language supports a stepwise design methodology and can serve as a rich and formal means of communication between the domain experts and the software engineers responsible for the system's design. The verification technique we provide would aid the engineers in checking that their design indeed meets the concrete requirements set by the MODs defined by the domain experts.

### 7.5 Synthesis and unsatisfiable cores

The most important future research we consider relates to a synthesis problem: given an MOD specification *MS*, find *cd* such that $cd \models MS$, if any. That is, we aim to develop an algorithm that takes as input a set of multi-modal object diagrams, made of positive and negative examples as well as positive and negative invariants, and outputs a CD that satisfies it (or reports that such a CD does not exist!). Note that for a classical set of ODs without modalities, each specifying a positive example, this problem is trivial, but for a multi-modal set it is much harder (and interesting). Also note that in many cases, there will be many possible solutions to the synthesis problem; we may be interested in synthesizing a satisfying CD that is minimal with regard to some cost function (e.g., depth or breadth of inheritance tree).

In the case where a satisfying CD cannot be synthesized, we will be interested in the related problem of finding an unsatisfiable core: a minimal subset of the MOD specification that has no satisfying CD (note that there may be more than one unsatisfiable core). The computation of an unsatisfiable core is a well known problem for SAT solvers. Unsatisfiable cores are essential means for the debugging of MOD specifications.

We hope to present the results of this research on synthesis and unsatisfiable cores for MOD specifications in a future paper.

## 8 Related Work

We discuss related work in adding modalities to existing modeling languages, in specifying constraints on ODs and combining them and OCL, in using Alloy for the analysis of class diagrams, and in other analysis problems related to class diagrams.

The idea that system models should include not only positive examples but also negative examples and positive and negative invariants is not new. This idea has been presented and investigated before in the context of behavioral models, in particular in scenario-based specifications. For example, the language of live sequence charts (LSC) [5, 12] extends classical message sequence charts (MSC) with universal and existential modalities, allowing to specify scenarios that must happen, scenarios that may happen, and scenarios that should never happen. In other variants of MSC [34], negative scenarios are used as a means for requirements elicitation and refinement. As in the case of MODs, the addition of modalities to the modeling language at hand, in this case, message sequence charts, results in a more expressive and useful language. It also comes with a price, in the form of a computationally expensive analysis (see, e.g., synthesis from LSC [11, 13]). Scenario-based specifications notwithstanding, we are not aware of any other study that investigates the addition of modalities in the context of structural system models, as we do with the introduction of modal object diagrams in the present paper.

Constraint diagrams [17] are a visual notation for specifying invariant constraints on object-oriented models, which can be viewed as a generalization of instance (object) diagrams, partly inspired by Venn diagrams. One may consider constraint diagrams to be similar to MODs, as both express constraints on object-oriented models. However, the two languages are fundamentally different. Constraint diagrams have their own visual notation while MODs only extend existing visual or textual notation with stereotypes. The invariants of constraint diagrams can be compared to the invariants of OCL (see Sect. 7.3 for a discussion on the relationship between OCL and MOD). To the best of our knowledge, based on [17], constraint diagrams cannot specify examples and have no explicit support for negation.

An integration of object diagrams and OCL for the specification of object-oriented systems is part of the definition of UML/P (see [27] ch. 5.3). This work proposes the embedding of object diagrams into OCL to, e.g., define the context of invariants, describe pre- and post-conditions, or specify relations between object diagrams (e.g., implication). Furthermore, elements like attributes and association links in object diagrams can be accessed from within the OCL/P syntax, relating the ODs to a system state. The semantics of logical conjunctives like `&&`, `||`, `implies`, etc. between ODs are informally given in [27]. Thus, the modalities of MOD can be expressed as an OCL/P predicate referencing UML/P ODs. Moreover, the other direction, of embedding OCL expressions into object diagrams, is defined too. For example, the language permits the definition of OCL/P variables inside object diagrams, e.g., to enable parametrized ODs, similar to the extension mentioned in Sect. 6.2. Our work is to a great extent inspired by these ideas. The UML/P language of [27] is far more expressive than MOD, however, it has no supporting reasoning mechanism and implementation. MOD can be viewed as a variant of UML/P ODs and their combination with OCL.

Some previous works consider the use of Alloy for the analysis of CDs (see, e.g., [1, 30]). These works focus on the formal definition of the transformation of a single CD to an Alloy module at the level of a meta-model and on its implementation using a transformation language. Possible applications of the use of Alloy to analyze a given CD are not discussed in depth in these works. In contrast, the input for our transformation consists not only of a class diagram but also of an object diagram (or a set of object diagrams). Moreover, the transformation itself is different, as it follows a pragmatic approach: we are not suggesting a meta-model level framework for general transformations but instead focus on solving the concrete verification problem we have at hand. Defining and implementing our transformation using QVT or other transformation language such as ATL [16] is possible, but is outside the focus of our work.

A different use of Alloy is considered in [29], where the authors present a meta-model directed model completion feature, in the context of code completion support in editors of domain-specific modeling languages. Although the setup and motivation are very different than ours, this work is somewhat similar to our work, specifically in the way predicates are used to define partial models.

Some previous works consider various analysis problems related to CDs (see, e.g., [2, 10, 21, 31, 33]). These include the finite satisfiability problem, the consistency of UML models (with or without OCL constraints), the identification of implicit consequences etc. Some of these use a direct translation into SAT and provide experimental performance results [31]. Others use Description Logic (DL) as their underlying formalism [33]. Some works include no implementation but present theoretical results about the decidability and complexity of the problems at hand. In contrast, we introduce a modal extension to the OD language and consider the problem of verifying that a given CD models a multi-modal specification. We provide a solution, in a bounded scope, using a reduction to an Alloy module and its analysis with a SAT solver.

Finally, in another paper in this conference [20] we have defined CDDiff, a semantic differencing operator for CDs (used for semantic model comparison in the context of model evolution), and have implemented it using a translation to Alloy. However, the translation we use for CDDiff is very different than the one we use here. The input for CDDiff consists of two CDs and its output is an OD that represents an OM that is in the semantics of the first CD and not in the semantics of the second. The input for MOD verification is a CD and an OD. While in CDDiff, each of the two input CDs is represented using a predicate, here we use the input CD as a base and the input OD induces a predicate that constrains it.

## 9  Conclusion

We introduced modal object diagrams, as an expressive extension to the classical object diagrams language. Moreover, we have presented a verification technique that can be used to verify, in a bounded scope, whether a given class diagram satisfies a multi-modal object diagram specification. We discussed a stepwise design process, where domain experts and analysts provide MODs while software engineers are responsible for designing class diagrams that satisfy them. The extended language and the verification technique are fully implemented in a prototype Eclipse plug-in.

The tradeoff of formality and expressiveness vs. intuitiveness and ease of use is a major challenge in modeling languages design. We believe that MOD addresses this tradeoff well: it is expressive enough to be valuable in specifying structural requirements of object-oriented systems, yet it is also intuitive and simple enough to be attractive to engineers.

Finally, we considered the advantages and limitations of our work. As discussed in Sect. 7, future work includes the development of heuristics to improve the performance of our verification technique and allow it to scale, the embedding of a subset of OCL inside the MOD language in order to extend its expressive power, performing case studies that will evaluate the use of MOD in the design of real-world systems, and the investigation of the problem of synthesizing a CD from a multi-modal object diagram specification.

**Acknowledgments** We are grateful to Martin Schindler for defining the MontiCore language support for ODs and CDs. We thank Smadar Szekely and Guy Weiss for their expert advice on Eclipse plug-in development. We thank Mira Balaban, David Lo, and the anonymous reviewers for comments on a draft of this paper.

# References

1. K. Anastasakis, B. Bordbar, G. Georg, and I. Ray. On challenges of model transformation from UML to Alloy. *Software and Systems Modeling*, 9(1):69–86, 2010.
2. D. Berardi, D. Calvanese, and G. D. Giacomo. Reasoning on UML class diagrams. *Artif. Intell.*, 168(1-2):70–118, 2005.
3. M. Broy, M. V. Cengarle, H. Grönniger, and B. Rumpe. Definition of the System Model. In K. Lano, editor, *UML 2 Semantics and Applications*. Wiley, 2009.
4. M. V. Cengarle, H. Grönniger, and B. Rumpe. System Model Semantics of Class Diagrams. Informatik-Bericht 2008-05, Technische Universität Braunschweig, 2008.
5. W. Damm and D. Harel. LSCs: Breathing life into Message Sequence Charts. *Formal Methods in System Design*, 19(1):45–80, 2001.
6. Dresden OCL. http://www.reuseware.org/index.php/DresdenOCL. Accessed 4/2011.
7. Eclipse UML2 project. http://www.eclipse.org/modeling/mdt/?project=uml2. Accessed 4/2011.
8. A. Evans, R. B. France, K. Lano, and B. Rumpe. The UML as a Formal Modeling Notation. In J. Bézivin and P.-A. Muller, editors, *Proc. 1st Int. Work. on the Unified Modeling Language, Selected Papers*, volume 1618 of *LNCS*, pages 336–348. Springer, 1998.
9. FreeMarker. http://freemarker.org/. Accessed 4/2011.
10. M. Gogolla, M. Kuhlmann, and L. Hamann. Consistency, independence and consequences in UML and OCL models. In C. Dubois, editor, *TAP*, volume 5668 of *LNCS*, pages 90–104. Springer, 2009.
11. D. Harel and H. Kugler. Synthesizing state-based object systems from LSC specifications. *Int. J. Found. Comput. Sci.*, 13(1):5–51, 2002.
12. D. Harel and S. Maoz. Assert and negate revisited: Modal semantics for UML sequence diagrams. *Software and Systems Modeling (SoSyM)*, 7(2):237–252, 2008.

13. D. Harel, S. Maoz, and I. Segall. Some results on the expressive power and complexity of LSCs. In A. Avron, N. Dershowitz, and A. Rabinovich, editors, *Pillars of Computer Science*, volume 4800 of *LNCS*, pages 351–366. Springer, 2008.
14. IBM Rational Software Architect (RSA). http://www.ibm.com/developerworks/rational/products/rsa/. Accessed 4/2011.
15. D. Jackson. *Software Abstractions: Logic, Language, and Analysis*. MIT Press, 2006.
16. F. Jouault, F. Allilaire, J. Bézivin, and I. Kurtev. ATL: A model transformation tool. *Sci. Comput. Program.*, 72(1-2):31–39, 2008.
17. S. Kent. Constraint diagrams: Visualizing assertions in object-oriented models. In *OOPSLA*, pages 327–341, 1997.
18. H. Krahn, B. Rumpe, and S. Völkel. MontiCore: a framework for compositional development of domain specific languages. *International Journal on Software Tools for Technology Transfer (STTT)*, 12(5):353–372, 2010.
19. E. Kuss. Using Alloy Analyzer for automated consistency checks between UML/P class and object diagrams. Master's thesis, Software Engineering, RWTH Aachen, Germany, 2010. In German.
20. S. Maoz, J. O. Ringert, and B. Rumpe. CDDiff: Semantic differencing for class diagrams. In *Proc. 25th Euro. Conf. on Object Oriented Programming (ECOOP'11)*, 2011. To appear.
21. A. Maraee and M. Balaban. Efficient reasoning about finite satisfiability of UML class diagrams with constrained generalization sets. In D. H. Akehurst, R. Vogel, and R. F. Paige, editors, *ECMDA-FA*, volume 4530 of *LNCS*, pages 17–31. Springer, 2007.
22. MOD project materials. http://www.se-rwth.de/materials/mod/.
23. MontiCore project. http://www.monticore.org/.
24. O. Nierstrasz. Ten things I hate about object-oriented programming. *Journal of Object Technology*, 9(5), Sept. 2010. (editorial [Banquet speech given at ECOOP 2010. Maribor, June 24, 2010]).
25. OMG (Object Management Group). Object Constraint Language (OCL). http://www.omg.org/spec/OCL/2.2/. Accessed 4/2011.
26. Poseidon for UML. http://www.gentleware.com/. Accessed 4/2011.
27. B. Rumpe. *Modellierung mit UML*. Springer, 2004.
28. SAT4J project. http://www.sat4j.org/. Accessed 4/2011.
29. S. Sen, B. Baudry, and H. Vangheluwe. Towards domain-specific model editors with automatic model completion. *Simulation*, 86(2):109–126, 2010.
30. S. M. A. Shah, K. Anastasakis, and B. Bordbar. From UML to Alloy and back again. In S. Ghosh, editor, *MoDELS Workshops*, volume 6002 of *LNCS*, pages 158–171. Springer, 2009.
31. M. Soeken, R. Wille, M. Kuhlmann, M. Gogolla, and R. Drechsler. Verifying UML/OCL models using Boolean satisfiability. In *DATE*, pages 1341–1344. IEEE, 2010.
32. Sparx Systems Enterprise Architect. http://www.sparxsystems.com/. Accessed 4/2011.
33. R. V. D. Straeten, T. Mens, J. Simmonds, and V. Jonckers. Using description logic to maintain consistency between UML models. In P. Stevens, J. Whittle, and G. Booch, editors, *Proc. 6th Int. Conf. on The Unified Modeling Language*, volume 2863 of *LNCS*, pages 326–340. Springer, 2003.
34. S. Uchitel, J. Kramer, and J. Magee. Negative scenarios for implied scenario elicitation. In *SIGSOFT FSE*, pages 109–118. ACM, 2002.